\documentclass[10pt, twocolumn, pre, aps, superscriptaddress, showpacs]{revtex4-1}
\usepackage{amsmath, graphicx, subfigure, tikz}
\begin{document}
    \title{Devil's Staircase Continuum in the Chiral Clock Spin Glass with Competing Ferromagnetic-Antiferromagnetic and Left-Right Chiral Interactions}
    \author{Tolga \c{C}a\u{g}lar}
    \affiliation{Faculty of Engineering and Natural Sciences, Sabanc\i\ University, Tuzla, Istanbul 34956, Turkey}
    \author{A. Nihat Berker}
    \affiliation{Faculty of Engineering and Natural Sciences, Sabanc\i\ University, Tuzla, Istanbul 34956, Turkey}
\affiliation{Faculty of Engineering and Natural Sciences, Kadir Has
University, Cibali, Istanbul 34083, Turkey}
    \affiliation{Department of Physics, Massachusetts Institute of Technology, Cambridge, Massachusetts 02139, USA}



\begin{abstract}
The chiral clock spin-glass model with $q=5$ states, with both
competing ferromagnetic-antiferromagnetic and left-right chiral
frustrations, is studied in $d=3$ spatial dimensions by
renormalization-group theory.  The global phase diagram is
calculated in temperature, antiferromagnetic bond concentration $p$,
random chirality strength, and right-chirality concentration $c$.
The system has a ferromagnetic phase, a multitude of different
chiral phases, a chiral spin-glass phase, and a critical
(algebraically) ordered phase. The ferromagnetic and chiral phases
accumulate at the disordered phase boundary and form a spectrum of
devil's staircases, where different ordered phases
characteristically intercede at all scales of phase-diagram space.
Shallow and deep reentrances of the disordered phase, bordered by
fragments of regular and temperature-inverted devil's staircases,
are seen. The extremely rich phase diagrams are presented as
continuously and qualitatively changing videos.
\end{abstract}
\maketitle

\section{Introduction}
The presence of chiral interactions, motivated by experimental
systems \cite{Ostlund, Kardar, Huse, Huse2, Caflisch}, can result in
extremely rich phase transition phenomena in otherwise simple
systems \cite{Caglar}.  In this respect, we study here a $q=5$ state
clock spin-glass model in $d=3$ spatial dimensions, using
renormalization-group theory.  Our system has both competing
ferromagnetic and antiferromagnetic interactions, as in the usually
studied spin-glass models \cite{NishimoriBook}, and competing
left-chiral and right-chiral interactions \cite{Caglar}. We have
studied $q=5$ states, because odd number of states have built-in
entropy for antiferromagnetic interactions, even without quenched
randomness and frustration.\cite{Ilker3}

The global phase diagram is calculated in temperature,
antiferromagnetic bond concentration $p$, random chirality strength,
and right-chirality concentration $c$. We find an extremely rich
phase diagram, with a ferromagnetic phase, a multitude of different
chiral phases, a chiral spin-glass phase, and a critical
(algebraically) ordered phase
\cite{BerkerKadanoff1,BerkerKadanoff2}. The ferromagnetic and chiral
phases accumulate at the disordered phase boundary and form a
devil's staircases \cite{Bak,Fukuda}, where different ordered phases
characteristically intercede at all scales of phase-diagram space.
In fact, a continuum of devil's staircases is found. Shallow and
deep reentrances of the disordered phase, bordered by fragments of
regular and temperature-inverted devil's staircases, are seen. The
extremely rich phase diagrams are presented as continuously and
qualitatively changing videos \cite{SupMat}.

\section{The $q-$state Chiral Clock\\
Double Spin Glass} The \textbf{$q-$state clock spin glass} is
composed of unit spins that are confined to a plane and that can
only point along $q$ angularly equidistant directions, with
Hamiltonian
\begin{equation}
-\beta {\cal H} = \sum_{\left<ij\right>} J_{ij} \vec{s}_i.\vec{s}_j
= \sum_{\left<ij\right>} J_{ij}\cos\theta_{ij}, \label{eq:qclockBH}
\end{equation}
where $\beta = 1/k_BT$, $\theta_{ij} = \theta_i - \theta_j$, at each
site $i$ the spin angle $\theta_i$ takes on the values
$(2\pi/q)\sigma_i$ with $\sigma_i=0,1,2,\ldots,(q-1)$, and
$\left<ij\right>$ denotes that the sum runs over all
nearest-neighbor pairs of sites. As a
ferromagnetic-antiferromagnetic spin-glass system
\cite{NishimoriBook}, the bond strengths $J_{ij}$, with quenched
(frozen) ferromagnetic-antiferromagnetic randomness, are $+J > 0$
(ferromagnetic) with probability $1-p$ and $-J$ (antiferromagnetic)
with probability $p$, with $0 \leq p \leq 1$. Thus, the
ferromagnetic and antiferromagnetic interactions locally compete in
frustration centers.  Recent studies on
ferromagnetic-antiferromagnetic clock spin glasses are in Refs.
\cite{Ilker1,Ilker3,Lupo}.

In the \textbf{$q-$state chiral clock double spin glass} introduced
here, frustration also occurs via randomly frozen left or right
chirality \cite{Caglar}.  The Hamiltonian in Eq. (1) is generalized
to random local chirality,
\begin{equation}
\label{eq:qccsgBH} -\beta {\cal H} = \sum_{\left<ij\right>} [
J_{ij}\cos\theta_{ij} + \Delta \, \delta(\theta_{ij} + \eta_{ij}
\frac{2\pi}{q})].
\end{equation}
In a cubic lattice, the $x,y,$ or $z$ coordinates increase as sites
along the respective coordinate direction are considered.
Bond-moving as in Fig. 1(a) is done transversely to the bond
directions, so that this sequencing is respected. Equivalently, in
the corresponding hierarchical lattice, one can always define a
direction along the connectivity, for example from left to right in
Fig. 1(b), and assign consecutive increasing number labels to the
sites.  In Eq. (2), for each pair of nearest-neighbor sites
$\left<ij\right>$ the numerical site label $j$ is ahead of $i$,
frozen (quenched) $\eta_{ij} = 1$ (left chirality) or $-1$ (right
chirality), and the delta function $\delta(x)=1\,(0)$ for $x=0\,
(x\neq 0)$. The overall concentrations of left and right chirality
are respectively $1-c$ and $c$, with $0 \leq c \leq 1$. The strength
of the random chiral interaction is $\Delta/J$, with temperature
divided out.  With no loss of generality, we take $\Delta \geq 0$.
Thus, the system is chiral for $\Delta > 0$, chiral-symmetric for
$c=0.5$, chiral-symmetry-broken for $c\neq0.5$. The global phase
diagram is in terms of temperature $J^{-1}$, antiferromagnetic bond
concentration $p$, random chirality strength $\Delta / J$, and
chiral symmetry-breaking concentration $c$.

\section{Renormalization-Group Method: Migdal-Kadanoff Approximation and Exact Hierarchical Lattice Solution}

We solve the chiral clock double spin-glass model with $q=5$ states
by renormalization-group theory, in $d=3$ spatial dimensions, with
length rescaling factor $b=3$.  We use $b=3$, as in previous
position-space renormalization-group calculations of spin-glass
systems, because it treats ferromagnetism and antiferromagnetism on
equal footing. Our solution is, simultaneously, the Migdal-Kadanoff
approximation \cite{Migdal,Kadanoff} for the cubic lattice and the
exact solution
\cite{BerkerOstlund,Kaufman1,Kaufman2,McKay,Hinczewski1} for the
$d=3$ hierarchical lattice based on the repeated self-imbedding of
leftmost graph of Fig. 1(b). Fig. 1(a) shows the Migdal-Kadanoff
approximate renormalization-group transformation for the cubic
lattice, composed of the bond-moving followed by decimation steps.
Fig. 1(b) shows the exact renormalization-group transformation for
the hierarchical lattice.  The two procedures yield identical
recursion relations.

Exact calculations on hierarchical lattices are also currently
widely used on a variety of statistical mechanics
problems.\cite{Timonin,Derrida,Thorpe,Efrat,Monthus2,
Hasegawa,Lyra,Singh,Xu2014,Hirose1,Silva,Hotta,Boettcher1,Boettcher2,Hirose2,Boettcher3,Nandy}.
On the other hand, this approximation for the cubic lattice is an
uncontrolled approximation, as in fact are all renormalization-group
theory calculations in $d=3$ and all mean-field theory calculations.
However, as noted before \cite{Yunus}, the local summation in
position-space technique used here has been qualitatively,
near-quantitatively, and predictively successful in a large variety
of problems, such as arbitrary spin-$s$ Ising models
\cite{BerkerSpinS}, global Blume-Emery-Griffiths model
\cite{BerkerWortis}, first- and second-order Potts transitions
\cite{NienhuisPotts,AndelmanBerker}, antiferromagnetic Potts
critical phases \cite{BerkerKadanoff1,BerkerKadanoff2}, ordering
\cite{BerkerPLG} and superfluidity \cite{BerkerNelson} on surfaces,
multiply reentrant liquid crystal phases \cite{Indekeu,Garland},
chaotic spin glasses \cite{McKayChaos}, random-field
\cite{Machta,FalicovRField} and random-temperature
\cite{HuiBerker,HuiBerkerE} magnets including the remarkably small
$d=3$ magnetization critical exponent $\beta$ of the random-field
Ising model, and high-temperature superconductors
\cite{HincewskiSuperc}.

\begin{figure}[ht!]
\centering
\includegraphics[scale=1.0]{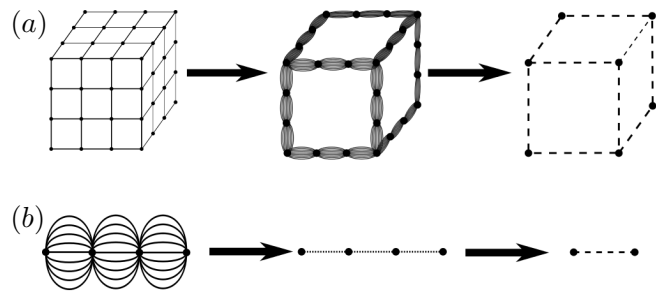}
\caption{(a) The Migdal-Kadanoff approximate renormalization-group
transformation for the cubic lattice, composed of the bond-moving
followed by decimation steps, with the length rescaling factor
$b=3$. The corresponding hierarchical lattice is obtained by the
repeated self-imbedding of the leftmost graph in (b). (b) The exact
renormalization-group transformation for this $d=3$ hierarchical
lattice. The two procedures yield identical recursion relations.}
\label{fig:rg_scheme}
\end{figure}

\begin{figure*}[ht!]
\centering
\includegraphics[scale=1]{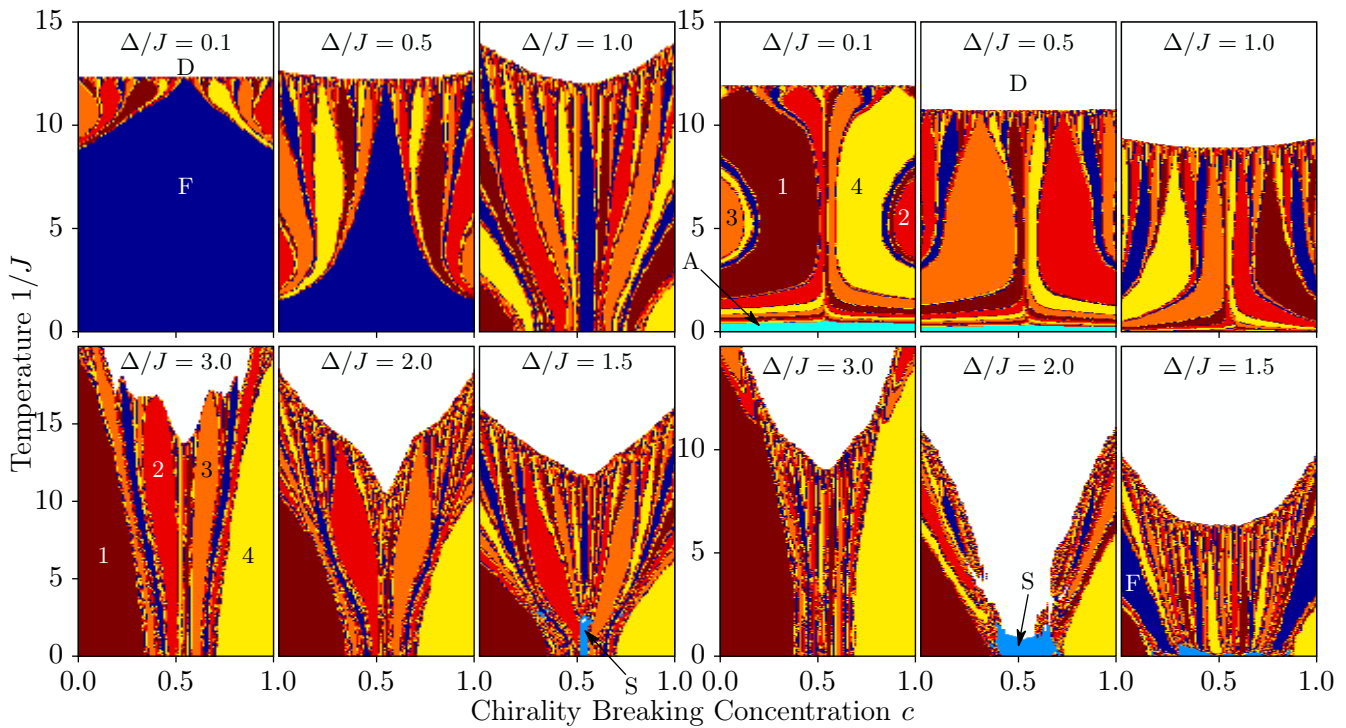}
\caption{(Color online) Calculated sequence of phase diagrams for
the ferromagnetic $(p=0)$, on the left side of the figure, and
antiferromagnetic $(p=1)$, on the right side, systems with quenched
random left- and right-chiral interactions. The horizontal axis $c$
is the concentration of right-chiral interactions. Phase diagrams
for different random chirality strengths $\Delta / J$ are shown. The
system exhibits ferromagnetic (F), a multitude of different chiral,
and spin-glass (S) ordered phases. On some of the chiral phases, the
$\delta$ multiplicity of the asymptotically dominant interaction is
indicated. The ferromagnetic and chiral phases accumulate as
different devil's staircases at their boundary with the disordered
(D) phase. The antiferromagnetic system also exhibits an
algebraically ordered (A) phase. The full richness of the continuum
of widely varying devil's staircase phase diagrams can also be seen
in video form, four of which are accessible as Supplemental Material
\cite{SupMat}. These four videos are also accessible at http://
web.mit.edu/physics/berker/temperatureDeltac0scanp.avi,
web.mit.edu/physics/berker/temperatureDeltac05scanp.avi,
web.mit.edu/physics/berker/temperaturecp1scanDelta.avi,
web.mit.edu/physics/berker/temperaturecp0scanDelta.avi}
\end{figure*}

\begin{figure*}[ht!]
\centering
\includegraphics[scale=1]{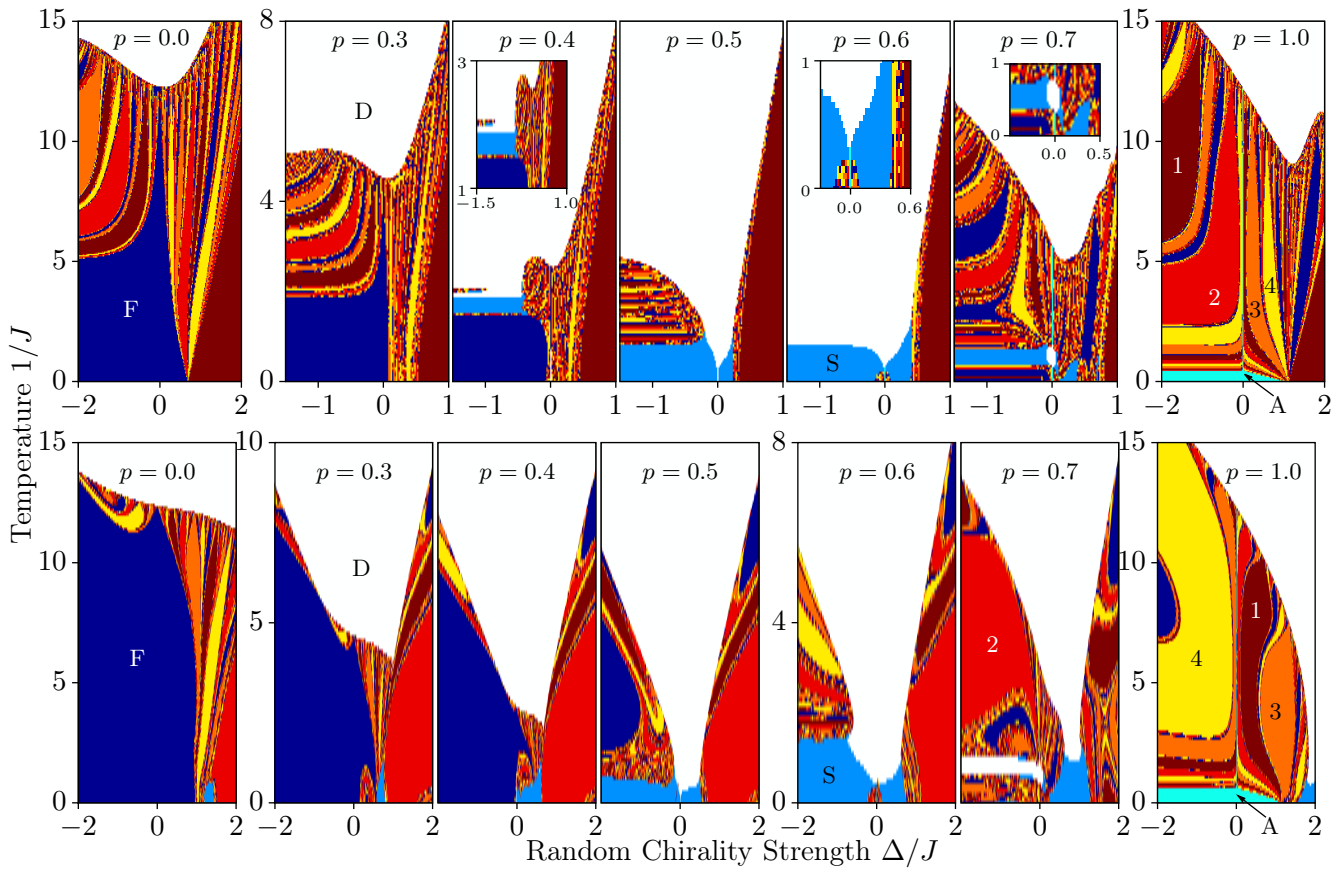}
\caption{(Color online) Calculated sequence of phase diagrams for
the left-chiral $(c=0)$, on the upper side of the figure, and
quenched random left- and right-chiral $(c=0.5)$, on the lower side,
systems with quenched random ferromagnetic and antiferromagnetic
interactions. The horizontal axis is the random chirality strength
$\Delta / J$. The consecutive phase diagrams are for different
concentrations of antiferromagnetic interactions $p$. The system
exhibits ferromagnetic (F), a multitude of different chiral, and
spin-glass (S), and critical (algebraically) ordered (A) phases. On
some of the chiral phases, the $\delta$ multiplicity of the
asymptotically dominant interaction is indicated. The ferromagnetic
and chiral phases accumulate as different devil's staircases at
their boundary with the disordered (D) phase. Note shallow and deep
reentrances of the disordered phase at $p=0.4$ and $p=0.7$,
respectively, surrounded by regular and temperature-inverted devil's
staircases. The full richness of the continuum of widely varying
devil's staircase phase diagrams can also be seen in video form,
four of which are accessible as Supplemental Material \cite{SupMat}.
These four videos are also accessible at http://
web.mit.edu/physics/berker/temperatureDeltac0scanp.avi,
web.mit.edu/physics/berker/temperatureDeltac05scanp.avi,
web.mit.edu/physics/berker/temperaturecp1scanDelta.avi,
web.mit.edu/physics/berker/temperaturecp0scanDelta.avi}
\end{figure*}

\begin{figure*}[ht!]
\centering
\includegraphics[scale=1]{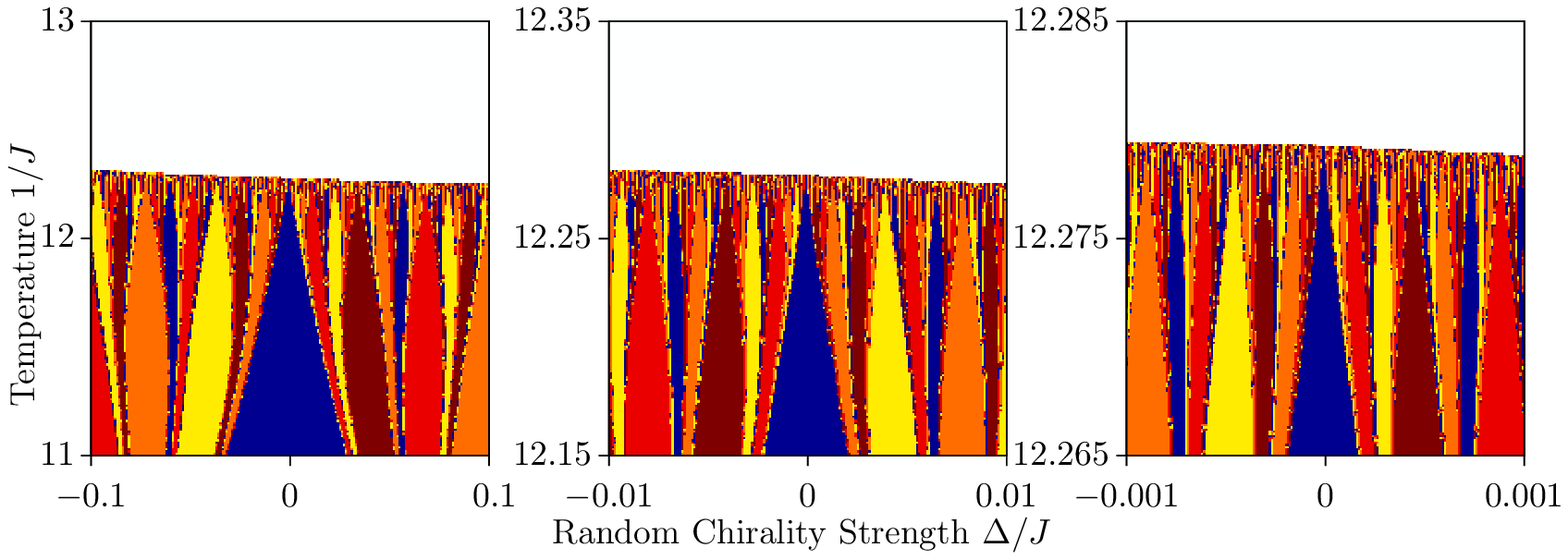}
\caption{(Color online) The phase diagram cross-section in the upper
left of Fig. 3, with a calculated 10-fold zoom and with 100-fold
zoom. The devil's staircase structure appears at each zoom level.}
\end{figure*}

Under the renormalization-group transformation described below, the
Hamiltonian of Eq. (2) maps onto the more general form
\begin{equation}
-\beta {\cal H} = \sum_{\left<ij\right>} V_{ij}(\theta_{ij}),
\label{eq:renhamil}
\end{equation}
where $\theta_{ij} = \theta_i - \theta_j$ can take $q$ different
values, so that for each pair $<ij>$ of nearest-neighbor sites,
there are q different interaction constants
\begin{multline}
\{V_{ij}(\theta_{ij})\} = \\
\{V_{ij}(0),V_{ij}(\delta),V_{ij}(2\delta),V_{ij}(3\delta),V_{ij}(4\delta)\}
\equiv \textbf{V}_{ij} ,
\end{multline}
which are in general different at each locality (quenched
randomness). Here, $\delta \equiv 2\pi/5$ is the angle between
consecutive clock-spin directions. The largest element of
$\{V_{ij}(\theta_{ij})\}$ at each locality $<ij>$ is set to zero, by
subtracting the same constant $G$ from all $q$ interaction
constants, with no effect on the physics; thus, the $q-1$ other
interaction constants are negative.

The local renormalization-group transformation is achieved by the
sequence, shown in Fig. 1, of bond movings
\begin{equation}
\widetilde{V}_{ij}(\theta_{ij})-\widetilde{G} =
\sum_{k=1}^{b^{d-1}}V_{ij}^{(k)}(\theta_{ij}), \label{eq:bondmove}
\end{equation}
and decimations
\begin{equation}
e^{V_{14}'(\theta_{14}) - G} =
\sum_{\theta_2,\theta_3}e^{\widetilde{V}_{12}(\theta_{12}) +
\widetilde{V}_{23}(\theta_{23}) + \widetilde{V}_{34}(\theta_{34})},
\label{eq:decimation}
\end{equation}
where $\widetilde{G}$ and $G$ are the subtractive constants
mentioned above, and prime marks the interaction of the renormalized
system.

The starting double-bimodal quenched probability distribution of the
interactions, characterized by $p$ and $c$ as described above, is
not conserved under rescaling. The renormalized quenched probability
distribution of the interactions is obtained by the convolution
\cite{Andelman}
\begin{equation}
P'(\textbf{V}'_{i'j'}) =
\int{\left\{\prod_{ij}^{i'j'}d\textbf{V}_{ij}P(\textbf{V}_{ij})\right\}\delta(\textbf{V}'_{i'j'}
- \textbf{R}(\left\{\textbf{V}_{ij}\right\}))}, \label{eq:conv}
\end{equation}
where $\textbf{V}_{ij} \equiv \{V_{ij}(\theta_{ij})\}$ as in Eq.
(4), $\textbf{R}(\{\textbf{V}_{ij}\})$ represents the bond moving
and bond decimation given in Eqs. \eqref{eq:bondmove} and
\eqref{eq:decimation}, and primes refer to the renormalized system.
Similar previous studies, on other spin-glass systems, are in Refs.
\cite{Gingras1,Migliorini,Gingras2,Heisenberg,Guven,Ohzeki,Ilker1,Ilker2,Ilker3,Demirtas}.
For numerical practicality the bond moving and decimation of Eqs.
(5) and (6) are achieved by a sequential pairwise combination of
interactions, each pairwise combination leading to an intermediate
probability distribution resulting from a pairwise convolution as in
Eq. (7).

We effect this procedure numerically, first starting with the
initial double delta distribution of Eq. (2) giving 4 possible
interactions quenched randomly distributed throughout the system,
and generating $1000$ interactions that embody the quenched
probability distribution resulting from the pairwise combination.
Each of the generated $1000$ interactions is described by $q$
interaction constants, as explained above [Eq. (4)]. At each
subsequent pairwise convolution as in Eq. (7), $1000$ randomly
chosen pairs, representing quenched random neighbors in the lattice,
are matched by (5) or (6), and a new set of $1000$ interactions is
produced. As a control, we have also calculated phase diagrams given
below using $1500$ interactions and the phase diagrams did not
change.

Our calculation simply consists in following the recursion
relations, Eqs.(5-7) to the various fixed points and thereby mapping
the initial conditions that are the basins of attraction of the
various fixed points.  This map is the phase diagram:  The different
thermodynamic phases of the system are identified by the different
asymptotic renormalization-group flows of the quenched probability
distribution $P(\textbf{V}_{ij})$.  Two renormalization-group
trajectories starting at each side of a phase boundary point diverge
from each other, flowing towards the phase sinks (completely stable
fixed points) of their respective phases.  Thus, the phase boundary
point between two phases is readily obtained to the accuracy of the
figures.  We are therefore able to calculate the global phase
diagram of the chiral clock double spin-glass model.\\

\section{Global Phase Diagram of the $q=5$\\
 State Chiral Clock Double Spin Glass}

The global phase diagram of the $q=5$ state chiral clock double
spin-glass model in $d=3$ spatial dimensions, in temperature
$J^{-1}$, antiferromagnetic bond concentration $p$, random chirality
strength $\Delta / J$, and right-chirality concentration $c$, is a
four-dimensional object, so that only the cross-sections of the
global phase diagram are exhibited.

Figs. 2 show the calculated sequence of phase diagrams for the
ferromagnetic $(p=0)$, on the left side of the figure, and
antiferromagnetic $(p=1)$, on the right side, systems with quenched
random left- and right-chiral interactions. The horizontal axis $c$
is the concentration of right-chiral interactions. Phase diagrams
for different random chirality strengths $\Delta / J$ are shown. The
system exhibits ferromagnetic (F), a multitude of different chiral,
and spin-glass (S) ordered phases. The antiferromagnetic system also
shows an algebraically (A) ordered (critical) phase, in which every
point is a critical point with divergent correlation length
\cite{BerkerKadanoff1,BerkerKadanoff2}. In all cases, the
ferromagnetic and different chiral phases accumulate as different
devil's staircases \cite{Bak,Fukuda} at their boundary with the
disordered (D) phase.  The definition of the devil's staircase is
that this accumulation is seen at every expanded scale of the phase
diagram variables.  This accumulation at every expanded phase
diagram scale is indeed revealed from our calculations, as seen
further below.

Figs. 3 show the calculated sequence of phase diagrams for the
left-chiral $(c=0)$, on the upper side, and quenched random left-
and right-chiral $(c=0.5)$, on the lower side, system with in both
cases quenched random ferromagnetic and antiferromagnetic
interactions. The horizontal axis is the random chirality strength
$\Delta / J$. The consecutive phase diagrams are for different
concentrations of antiferromagnetic interactions $p$. The system
exhibits ferromagnetic (F), a multitude of different chiral,
spin-glass (S), and algebraically ordered (A) phases. The
ferromagnetic and different chiral phases accumulate as different
devil's staircases \cite{Bak,Fukuda} at their boundary with the
disordered (D) phase. Note shallow and deep reentrances of disorder
\cite{Cladis,Hardouin,Garland,Netz,Kumari} at $p=0.4$ and $p=0.7$,
respectively, surrounded by regular and temperature-inverted devil's
staircases.

Fig. 4 shows the phase diagram cross-section in the upper left of
Fig. 3, with a calculated 10-fold zoom and with 100-fold zoom.  The
devil's staircase structure appears at each zoom level.

The full richness of the continuum of widely varying devil's
staircase phase diagrams can best be seen in video form, four of
which are accessible as Supplemental Material \cite{SupMat}.  These
four videos are also accessible at http://
web.mit.edu/physics/berker/temperatureDeltac0scanp.avi,
web.mit.edu/physics/berker/temperatureDeltac05scanp.avi,
web.mit.edu/physics/berker/temperaturecp1scanDelta.avi,
web.mit.edu/physics/berker/temperaturecp0scanDelta.avi. These videos
effectively exhibit a very large number of calculated phase diagram
cross-sections.

\section{Entire-Phase Criticality, Differentiated Chaos in the Spin-Glass and at its Boundary }

The renormalization-group mechanism for the algebraically ordered
(critical) phase is that, all renormalization-group trajectories
originating inside this phase flow to a completely stable fixed
point (sink) that occurs at finite temperature (finite coupling
strength).\cite{BerkerKadanoff1,BerkerKadanoff2,Saleur,Jacobsen1,Jacobsen2,Ikhlef,Jacobsen3,Jacobsen4,Jacobsen5,Bondesan,Jacobsen6}
In all other ordered phases, the trajectories flow to strong
(infinite) coupling.

In the ferromagnetic phase, the interaction $V_{ij}(0)$ becomes
asymptotically dominant.  In the chiral phases, in the
renormalization-group trajectories, one of the chiral interactions
from the right-hand side of Eq. (4),
$\{V_{ij}(\delta),V_{ij}(2\delta),V_{ij}(3\delta),V_{ij}(4\delta)\}$,
becomes asymptotically dominant. However, in each of the separate
phases, it takes a characteristic number $n$ of
renormalization-group transformations, namely a length scale of
$3^n$, to reach the dominance of one chiral interaction. This
distinct number of iterations, namely scale changes, determines, by
tracing back to the periodic sequence in the original lattice, the
pitch of the chiral phase in the original unrenormalized system.
Thus, the chiral phases in the original unrenormalized system, with
distinct chiral pitches, are distinct phases. After the dominance of
one chiral interaction, the renormalization-group trajectory follows
the periodic sequence $V_{ij}(\delta) \rightarrow V_{ij}(3\delta)
\rightarrow V_{ij}(4\delta) \rightarrow V_{ij}(2\delta) \rightarrow
V_{ij}(\delta)$ resulting from matching $q=5$ and $b=3$.

Our calculation is exact for the hierarchical lattice pictured in
Fig. 1(b) therefore for which the phase diagrams in Fig. 2 and 3 are
exactly applicable.  However, our calculation is approximate for the
cubic lattice, as pictured in Fig. 1(a).  Thus, one could speculate
that in the cubic lattice, the multitude of chiral phases would
appear as a single chiral phase with a continuously varying pitch:
Fig. 5 shows all the chiral phases merged into a single phase.  It
is seen that a quite unusual phase diagram still appears, with the
interlacing of the ferromagnetic phase with the chiral phase,
throughout the bulk of the phase region.

\begin{figure*}[ht!]
\centering
\includegraphics[scale=1]{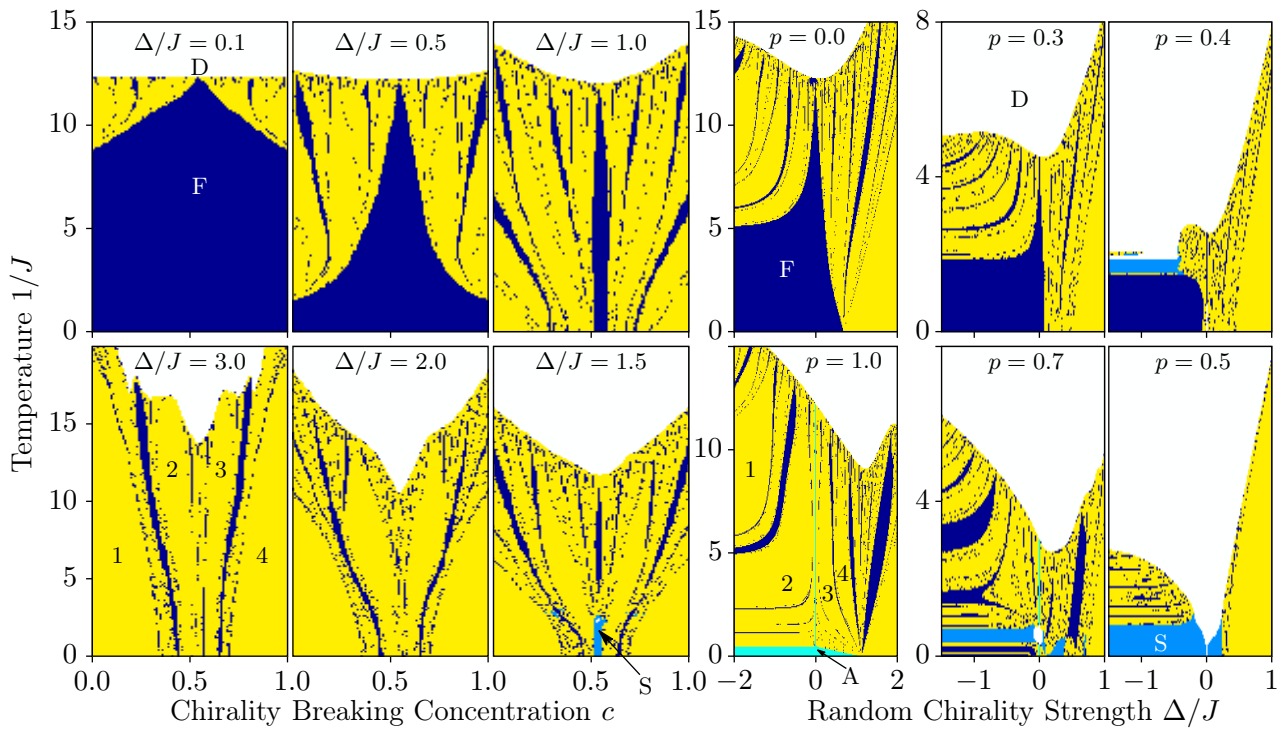}
\caption{(Color online) Our calculation is exact for the
hierarchical lattice pictured in Fig. 1(b), therefore for which the
phase diagrams in Fig. 2 and 3 are exactly applicable. However, our
calculation is approximate for the cubic lattice, as pictured in
Fig. 1(a). Thus, it could be speculated that in the cubic lattice,
the multitude of chiral phases would appear as a single chiral phase
with a continuously varying pitch: This Fig. 5 shows all the chiral
phases merged into a single phase.  It is seen that a quite unusual
phase diagram still appears, with the interlacing of the
ferromagnetic phase with the chiral phase, throughout the bulk of
the phase region. The left side of this figure is derived from the
left portion of Fig. 2; the right side is derived from the top
portion of Fig. 3}
\end{figure*}

\begin{figure}[ht!]
\centering
\includegraphics[scale=1]{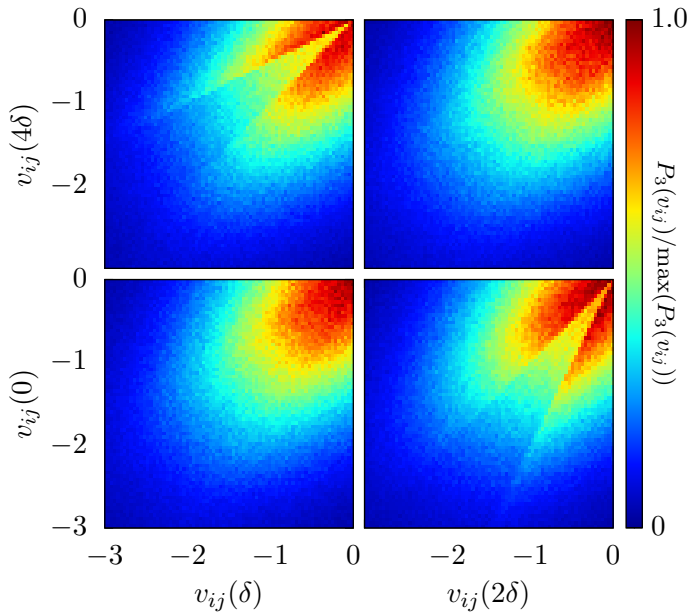}
\caption{(Color online) Asymptotic fixed distribution of the
spin-glass phase. The part of the fixed distribution,
$P_3(\textbf{V}_{ij})$ for the interactions $\textbf{V}_{ij}$ in
which $V_{ij}(3\delta)$ is maximum and therefore 0 (and the other
four interactions are negative) is shown in this figure, with
$v_{ij}(\sigma\delta) = V_{ij}(\sigma\delta)
/<|V_{ij}(\sigma\delta)|>$. The projections of
$P_3(\textbf{V}_{ij})$ onto two of its four arguments are shown in
each panel of this figure. The other four
$P_\sigma(\textbf{V}_{ij})$ have the same fixed distribution.  Thus,
chirality is broken locally, but not globally.}
\end{figure}

\begin{figure}[ht!]
\centering
\includegraphics[scale=1]{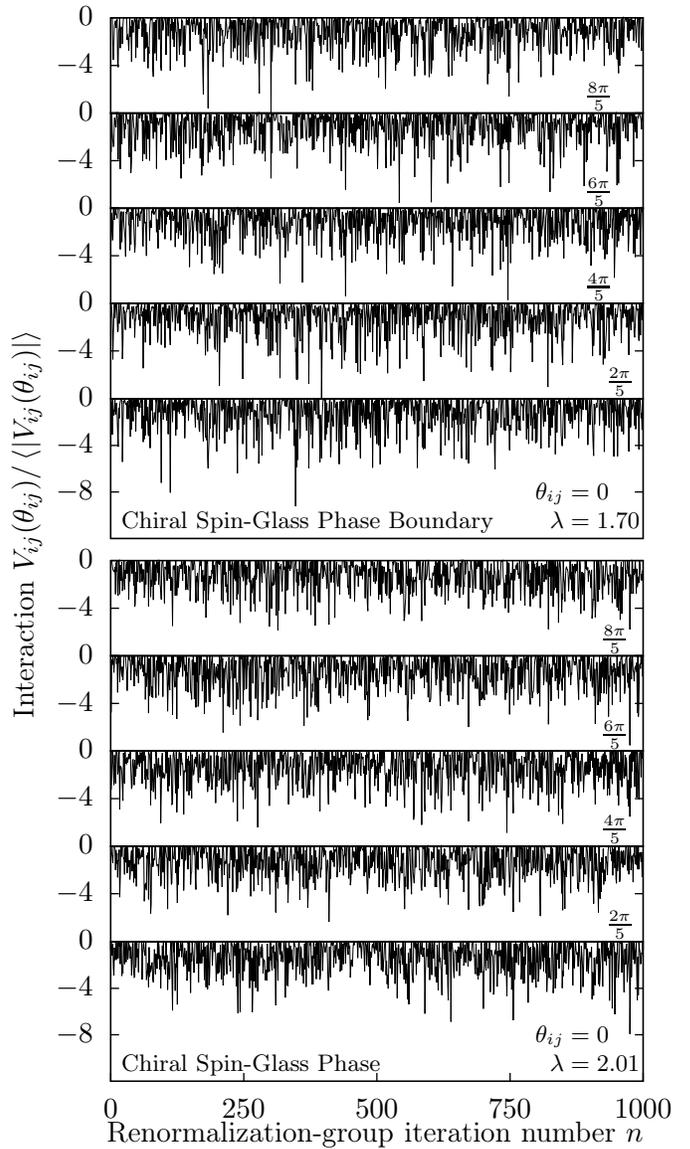}
\caption{Chaotic renormalization-group trajectories of the
spin-glass phase (bottom) and of the phase boundary between the
spin-glass and disordered phases (top). The five interactions
$V_{ij}(0),V_{ij}(\delta),V_{ij}(2\delta),V_{ij}(3\delta),V_{ij}(4\delta)$
at a given location $<ij>$, under consecutive renormalization-group
transformations, are shown. The $\theta_{ij} = \sigma \delta$
angular value of each interaction $V_{ij}(\theta_{ij})$ is indicated
in the figure panels. Bottom panel: Inside the spin-glass phase. The
corresponding Lyapunov exponent is $\lambda = 2.01$ and the average
interaction diverges as $<|V|> \sim b^{y_R n}$, where $n$ is the
number of renormalization-group iterations and $y_R = 0.26$ is the
runaway exponent. Top panel: At the phase boundary between the
spin-glass and disordered phases. The corresponding Lyapunov
exponent is $\lambda = 1.70$ and the average non-zero interaction
remains fixed at $<V> = -0.99$. As indicated by the Lyapunov
exponents, chaos is stronger inside the spin-glass phase than at its
phase boundary.}
\end{figure}

The renormalization-group trajectories starting in the chiral
spin-glass phase, unlike those in the ferromagnetic or chiral
phases, do not have the asymptotic behavior where at any scale a
single potential V(theta) is dominant.  These trajectories of the
spin-glass phase asymptotically go to a strong-coupling fixed
probability distribution $P(\textbf{V}_{ij})$ which assigns non-zero
probabilities to a distribution of $\textbf{V}_{ij}$ values, with no
single $V_{ij}(\theta)$ being dominant. Projections of this
distribution (a function of five variables) are shown in Fig. 6.
This situation is a direct generalization of the asymptotic
trajectories of the $\pm J$ Ising spin-glass phase, where a fixed
probability distribution over positive and negative values of the
interaction $J$ is obtained, with no single value of $J$ being
dominant \cite{Ilker1}.

Since, at each locality, the largest interaction in
$\{V_{ij}(0),V_{ij}(\delta),V_{ij}(2\delta),V_{ij}(3\delta),V_{ij}(4\delta)\}$
is set to zero and the four other interactions are thus made
negative, by subtracting the same constant from all five
interactions without affecting the physics, the quenched probability
distribution $P(\textbf{V}_{ij})$, a function of five variables, is
actually composed of five functions $P_\sigma(\textbf{V}_{ij})$ of
four variables, each such function corresponding to one of the
interactions being zero and the other four, arguments of the
function, being negative. Fig. 6 shows one of the latter functions:
The part of the fixed distribution, $P_3(\textbf{V}_{ij})$, for the
interactions $\textbf{V}_{ij}$ in which $V_{ij}(3\delta)$ is maximum
and therefore 0 (and the other four interactions are negative) is
shown in this figure. The projections of $P_3(\textbf{V}_{ij})$ onto
two of its four arguments are shown in each panel of this figure.
The other four $P_\sigma(\textbf{V}_{ij})$ have the same fixed
distribution. Thus, chirality is broken locally, but not globally.

Another distinctive mechanism, that of chaos under scale change
\cite{McKayChaos, McKayChaos2,BerkerMcKay} or, equivalently, under
spatial translation \cite{Ilker1}, occurs within the spin-glass
phase and differently at the spin-glass phase boundary
\cite{Ilker1}, in systems with competing ferromagnetic and
antiferromagnetic interactions \cite {McKayChaos,
McKayChaos2,BerkerMcKay,Bray,Hartford,
Nifle1,Nifle2,Banavar,Frzakala1,Frzakala2,Sasaki,Lukic,Ledoussal,Rizzo,
Katzgraber,Yoshino,Pixley,Aspelmeier1,Aspelmeier2,Mora,Aral,Chen,Jorg,Lima,Katzgraber2,MMayor,ZZhu,Katzgraber3,Fernandez,Ilker1,Ilker2,MMayor2}
and, more recently, with competing left- and right-chiral
interactions \cite{Caglar}.  The physical hierarchical lattice that
we solve here is an infinite system, where 1000 quintuplets
$\{V_{ij}(0),V_{ij}(\delta),V_{ij}(2\delta),V_{ij}(3\delta),V_{ij}(4\delta)\}$
are randomly distributed over the lattice bond positions.  Thus, as
we can fix our attention to one lattice position and monitor how the
quintuplet at that position evolves under renormalization-group
transformation, as it merges with its neighbors through bond moving
[Eq. (5)] and decimation [Eq. (6)], and thereby calculate the
Lyapunov exponent \cite{Ilker1,Ilker2}, which when positive is the
measure of the strength of chaos.

Fig. 7 gives the asymptotic chaotic renormalization-group
trajectories of the spin-glass phase and, distinctly, of the phase
boundary between the spin-glass and disordered phases. The chaotic
trajectories found here are similar to those found in traditional
(Ising) spin-glasses \cite{Ilker1,Ilker2}, with of course different
Lyapunov exponents seen below. The five interactions
$V_{ij}(0),V_{ij}(\delta),V_{ij}(2\delta),V_{ij}(3\delta),V_{ij}(4\delta)$
at a given location $<ij>$, under consecutive renormalization-group
transformations, are shown in Fig. 7. As noted, chaos is measured by
the Lyapunov exponent \cite{Collet,Hilborn,Aral,Ilker1,Ilker2},
which we here generalize, by the matrix form, to our
multi-interaction case:
\begin{equation}
\lambda = \lim _{n\rightarrow\infty} \frac{1}{n} \ln \Big|{\cal E}
\Big( \prod_{k=0}^{n-1} \frac {d\textbf{v}_{k+1}}{d\textbf{v}_k}
\Big) \Big| ,
\end{equation}
where the function ${\cal E}(\textbf{M})$ gives the largest
eigenvalue of its matrix argument $\textbf{M}$ and the vector
$\textbf{v}_k$ is
\begin{equation}
\textbf{v}_k = \{
v_{ij}(0),v_{ij}(\delta),v_{ij}(2\delta),v_{ij}(3\delta),v_{ij}(4\delta)
\},
\end{equation}
with $v_{ij}(\sigma\delta) = V_{ij}(\sigma\delta)
/<|V_{ij}(\sigma\delta)|>$, at step $k$ of the renormalization-group
trajectory.  The product in Eq. (8) is to be taken within the
asymptotic chaotic band, which is renormalization-group stable or
unstable for the spin-glass phase or its boundary, respectively.
Thus, we throw out the first 100 renormalization-group iterations to
eliminate the transient points outside of, but leading to the
chaotic band. Subsequently, typically using 1,000
renormalization-group iterations in the product in Eq. (8) assures
the convergence of the Lyapunov exponent value $\lambda$, which is
thus accurate to the number of significant figures given. Spin-glass
chaos occurs for $\lambda > 0$ \cite{Aral} and the more positive
$\lambda$, the stronger is chaos, as seen for example in the
progressions in Figs. 6 and 7 of Ref. \cite{Ilker2}. In the
spin-glass phase of the currently studied system, the Lyapunov
exponent is $\lambda = 2.01$ and the average interaction diverges as
$<|V|> \sim b^{y_R n}$, where $n$ is the number of
renormalization-group iterations and $y_R = 0.26$ is the runaway
exponent. At the phase boundary between the spin-glass and
disordered phases, the Lyapunov exponent is $\lambda = 1.70$ and the
average non-zero interaction remains fixed at $<V> = -0.99 $. As
indicated by the Lyapunov exponents, chaos is stronger inside the
spin-glass phase than at its phase boundary.

\section{Conclusion}

It is thus seen that chirality and chiral quenched randomness
provides, in a simple model, remarkably rich phase transition
phenomena.  These include a multitude of chiral phases, a continuum
of widely varying devil's staircases, shallow and deep reentrances
of the disordered phase surrounded by regular and
temperature-inverted devil's staircases, a critical phase, and a
chiral spin-glass phase with chaotic rescaling behavior inside and
differently at its boundary. The widely varying continuum of devil's
staircase phase diagrams are best seen in video form, four of which
are accessible as Supplemental Material \cite{SupMat}. These four
videos are also accessible at http://
web.mit.edu/physics/berker/temperatureDeltac0scanp.avi,
web.mit.edu/physics/berker/temperatureDeltac05scanp.avi,
web.mit.edu/physics/berker/temperaturecp1scanDelta.avi,
web.mit.edu/physics/berker/temperaturecp0scanDelta.avi.  Finally,
the study of an even number of $q$ states, which do not have a
built-in entropy as mentioned above, should yield equally rich, but
qualitatively different phase diagrams.

\begin{acknowledgments}
Support by the Academy of Sciences of Turkey (T\"UBA) is gratefully
acknowledged.
\end{acknowledgments}


\begin{references}

\bibitem{Ostlund} S. Ostlund, Phys. Rev. {\bf24}, 398 (1981).
\bibitem{Kardar} M. Kardar and A. N. Berker, Phys. Rev. Lett. {\bf48}, 1552 (1982).
\bibitem{Huse} D. A. Huse and M. E. Fisher, Phys. Rev. Lett. {\bf49}, 793 (1982).
\bibitem{Huse2} D. A. Huse and M. E. Fisher, Phys. Rev. {\bf29}, 239 (1984).
\bibitem{Caflisch} R. G. Caflisch, A. N. Berker, and M. Kardar, Phys. Rev. B {\bf31}, 4527 (1985).
\bibitem{Caglar} T. \c{C}a\u{g}lar and A. N. Berker, Phys. Rev. E {\bf 94}, 032121 (2016).
\bibitem{Ilker3} E. Ilker and A. N. Berker, Phys. Rev. E {\bf 90}, 062112 (2014).
\bibitem{NishimoriBook} H. Nishimori, \textit{Statistical Physics of Spin Glasses and Information
Processing} (Oxford University Press, Oxford, 2001).
\bibitem{BerkerKadanoff1} A. N. Berker and L. P. Kadanoff, J. Phys. A {\bf 13}, L259 (1980).
\bibitem{BerkerKadanoff2} A. N. Berker and L. P. Kadanoff, J. Phys. A {\bf 13}, 3786 (1980).
\bibitem{Bak} P. Bak and R. Bruinsma, Phys. Rev. Lett. {\bf 49}, 249 (1982).
\bibitem{Fukuda} A. Fukuda, Y. Takanishi, T. Isozaki, K. Ishikawa,
and H. Takezoe, J. Mat. Chem. {\bf 4}, 997 (1994).

\bibitem{SupMat} See Supplemental Material at http://link.aps.org/supplemental/
10.1103/PhysRevE.95.042125 for extremely rich devil's staircase
phase diagrams presented as continuously and qualitatively changing
4 videos. These four videos are also accessible at http://
web.mit.edu/physics/berker/temperatureDeltac0scanp.avi,
web.mit.edu/physics/berker/temperatureDeltac05scanp.avi,
web.mit.edu/physics/berker/temperaturecp1scanDelta.avi,
web.mit.edu/physics/berker/temperaturecp0scanDelta.avi

\bibitem{Ilker1} E. Ilker and A. N. Berker, Phys. Rev. E {\bf 87}, 032124 (2013).
\bibitem{Lupo} C. Lupo and F. Ricci-Tersenghi, Phys. Rev. B {\bf 95}, 054433 (2017).

\bibitem{Migdal} A. A. Migdal, Zh. Eksp. Teor. Fiz. {\bf69}, 1457 (1975) [Sov. Phys. JETP {\bf42}, 743 (1976)].
\bibitem{Kadanoff} L. P. Kadanoff, Ann. Phys. (N.Y.) {\bf100}, 359 (1976).

\bibitem{BerkerOstlund} A. N. Berker and S. Ostlund, J. Phys. C {\bf 12}, 4961 (1979).
\bibitem{Kaufman1} R. B. Griffiths and M. Kaufman, Phys. Rev. B {\bf 26}, 5022R (1982).
\bibitem{Kaufman2} M. Kaufman and R. B. Griffiths, Phys. Rev. B {\bf 30}, 244 (1984).
\bibitem{McKay} S. R. McKay and A. N. Berker, Phys. Rev. B {\bf29}, 1315 (1984).
\bibitem{Hinczewski1} M. Hinczewski and A. N. Berker, Phys. Rev. E {\bf 73}, 066126 (2006).

\bibitem{Timonin} P. N. Timonin, Low Temp. Phys. {\bf 40}, 36 (2014).
\bibitem{Derrida} B. Derrida and G. Giacomin, J. Stat. Phys. {\bf 154}, 286 (2014).
\bibitem{Thorpe} M. F. Thorpe and R. B. Stinchcombe, Philos. Trans. Royal Soc. A - Math. Phys. Eng. Sciences {\bf 372}, 20120038 (2014).
\bibitem{Efrat} A. Efrat and M. Schwartz, Physica {\bf 414}, 137 (2014).
\bibitem{Monthus2} C. Monthus and T. Garel, Phys. Rev. B {\bf 89}, 184408 (2014).
\bibitem{Hasegawa} T. Nogawa and T. Hasegawa, Phys. Rev. E {\bf 89}, 042803 (2014).
\bibitem{Lyra} M. L. Lyra, F. A. B. F. de Moura, I. N. de Oliveira, and M. Serva, Phys. Rev. E {\bf 89}, 052133 (2014).
\bibitem{Singh} V. Singh and S. Boettcher, Phys. Rev. E {\bf 90}, 012117 (2014).
\bibitem{Xu2014} Y.-L. Xu, X. Zhang, Z.-Q. Liu, K. Xiang-Mu, and R. Ting-Qi, Eur. Phys. J. B {\bf 87}, 132 (2014).
\bibitem{Hirose1} Y. Hirose, A. Oguchi, and Y. Fukumoto, J. Phys. Soc. Japan {\bf 83}, 074716 (2014).
\bibitem{Silva} V. S. T. Silva, R. F. S. Andrade, and S. R. Salinas, Phys. Rev. E {\bf 90}, 052112 (2014).
\bibitem{Hotta} Y. Hotta, Phys. Rev. E {\bf 90}, 052821 (2014).
\bibitem{Boettcher1} S. Boettcher, S. Falkner, and R. Portugal, Phys. Rev. A {\bf 91} 052330 (2015).
\bibitem{Boettcher2} S. Boettcher and C. T. Brunson, Eur. Phys. Lett. {\bf 110}, 26005 (2015).
\bibitem{Hirose2} Y. Hirose, A. Ogushi, and Y. Fukumoto, J. Phys. Soc. Japan {\bf 84}, 104705 (2015).
\bibitem{Boettcher3} S. Boettcher and L. Shanshan, J. Phys. A {\bf 48}, 415001 (2015).
\bibitem{Nandy} A. Nandy and A. Chakrabarti, Phys. Lett. {\bf 379}, 43 (2015).

\bibitem{Yunus} \c{C}. Yunus, B. Renklio\u{g}lu, M. Keskin, and A. N. Berker, Phys. Rev. E {\bf 93}, 062113 (2016).
\bibitem{BerkerSpinS} A. N. Berker, Phys. Rev. B {\bf 12}, 2752 (1975).
\bibitem{BerkerWortis} A. N. Berker and M. Wortis, Phys. Rev. B {\bf 14}, 4946 (1976).
\bibitem{NienhuisPotts} B. Nienhuis, A. N. Berker, E. K. Riedel, and M. Schick, Phys. Rev. Let. {\bf 43}, 737 (1979).
\bibitem{AndelmanBerker} D. Andelman and A. N. Berker, J. Phys. A {\bf 14}, L91 (1981).
\bibitem{BerkerPLG} A. N. Berker, S. Ostlund, and F. A. Putnam, Phys. Rev. B {\bf 9}, 3650 (1978).
\bibitem{BerkerNelson} A. N. Berker and D. R. Nelson, Phys. Rev. B {\bf 19}, 2488 (19769).
\bibitem{Indekeu} J. O. Indekeu and A. N. Berker, Physica A {\bf140}, 368 (1986).
\bibitem{Garland} J. O. Indekeu, A. N. Berker, C. Chiang, and C. W. Garland, Phys. Rev. A {\bf35}, 1371 (1987).
\bibitem{McKayChaos} S. R. McKay, A. N. Berker, and S. Kirkpatrick, Phys. Rev. Lett. {\bf 48}, 767 (1982).
\bibitem{Machta} M. S. Cao and J. Machta, Phys. Rev. B {\bf 48}, 3177 (1993).
\bibitem{FalicovRField} A. Falicov, A. N. Berker, and S. R. McKay, Phys. Rev. B {\bf 51}, 8266 (1995).
\bibitem{HuiBerker} K. Hui and A. N. Berker, Phys. Rev. Lett. {\bf 62}, 2507 (1989).
\bibitem{HuiBerkerE} K. Hui and A. N. Berker, Phys. Rev. Lett. {\bf 63}, 2433 (1989).
\bibitem{HincewskiSuperc} M. Hinczewski and A. N. Berker, Phys. Rev. B {\bf 78}, 064507 (2008).

\bibitem{Andelman} D. Andelman and A. N. Berker, Phys. Rev. B {\bf29}, 2630 (1984).

\bibitem{Gingras1} M. J. P. Gingras and E. S. S{\o}rensen, Phys. Rev. B. {\bf 46}, 3441 (1992).
\bibitem{Migliorini} G. Migliorini and A. N. Berker, Phys. Rev. B. {\bf 57}, 426 (1998).
\bibitem{Gingras2} M. J. P. Gingras and E. S. S{\o}rensen, Phys. Rev. B. {\bf 57}, 10264 (1998).
\bibitem{Heisenberg} C. N. Kaplan and A. N. Berker, Phys. Rev. Lett. {\bf 100}, 027204 (2008).
\bibitem{Guven} C. G\"{u}ven, A. N. Berker, M. Hinczewski, and H. Nishimori, Phys. Rev. E {\bf 77}, 061110 (2008).
\bibitem{Ohzeki} M. Ohzeki, H. Nishimori, and A. N. Berker, Phys. Rev. E {\bf 77}, 061116 (2008).
\bibitem{Ilker2} E. Ilker and A. N. Berker, Phys. Rev. E {\bf 89}, 042139 (2014).
\bibitem{Demirtas} M. Demirta\c{s}, A. Tuncer, and A. N. Berker, Phys. Rev. E {\bf 92}, 022136 (2015).

\bibitem{Cladis} P. E. Cladis, Phys. Rev. Lett. {\bf35}, 48 (1975).
\bibitem{Hardouin} F. Hardouin, A. M. Levelut, M. F. Achard, and G. Sigaud, J. Chim. Phys. {\bf80}, 53 (1983).
\bibitem{Netz} R. R. Netz and A. N. Berker, Phys. Rev. Lett. {\bf68}, 333 (1992).
\bibitem{Kumari} S. Kumari and S. Singh, Phase Transitions {\bf88}, 1225 (2015).

\bibitem{Saleur} H. Saleur, Nucl. Phys. B {\bf 360}, 219 (1991).
\bibitem{Jacobsen1} J. L. Jacobsen, J. Salas, and A. D. Sokal, J. Stat. Phys. {\bf 119}, 1153 (2005).
\bibitem{Jacobsen2} J. L. Jacobsen and H. Saleur, Nucl. Phys. B {\bf 743}, 207 (2006).
\bibitem{Ikhlef} Y. Ikhlef, Mod. Phys. Lett. {\bf 25}, 291 (2011).
\bibitem{Jacobsen3} J. L. Jacobsen and C. R. Scullard, J. Phys. A {\bf 45}, 494003 (2012).
\bibitem{Jacobsen4} J. L. Jacobsen and J. Salas, Nucl. Phys. B {\bf 875}, 678 (2013).
\bibitem{Jacobsen5} C. R. Scullard and J. L. Jacobsen, J. Phys. A {\bf 49}, 125003 (2016).
\bibitem{Bondesan} R. Bondesan, S. Caracciolo, and A. Sportiello, J. Phys. A: Math. Theor. 50, 074003 (2017).
\bibitem{Jacobsen6} J. L. Jacobsen, J. Salas, and C. R. Scullard, arXiv:1702.02006 [cond-mat.stat-mech] (2017).

\bibitem{McKayChaos2} S. R. McKay, A. N. Berker, and S. Kirkpatrick, J. Appl. Phys. {\bf 53}, 7974 (1982).
\bibitem{BerkerMcKay} A. N. Berker and S. R. McKay, J. Stat. Phys. {\bf 36}, 787 (1984).

\bibitem{Bray} A. J. Bray and M. A. Moore, Phys. Rev. Lett. {\bf 58}, 57 (1987).
\bibitem{Hartford} E. J. Hartford and S. R. McKay, J. Appl. Phys. 70, 6068 (1991).
\bibitem{Nifle1} M. Nifle and H. J. Hilhorst, Phys. Rev. Lett. {\bf 68}, 2992 (1992).
\bibitem{Nifle2} M. Nifle and H. J. Hilhorst, Physica A {\bf 194}, 462 (1993).
\bibitem{Banavar} M. Cieplak, M. S. Li, and J. R. Banavar, Phys. Rev. B {\bf 47}, 5022 (1993).
\bibitem{Frzakala1} F. Krzakala, Europhys. Lett. {\bf 66}, 847 (2004).
\bibitem{Frzakala2} F. Krzakala and J. P. Bouchaud, Europhys. Lett. {\bf 72}, 472 (2005).
\bibitem{Sasaki} M. Sasaki, K. Hukushima, H. Yoshino, and H. Takayama, Phys. Rev. Lett. {\bf 95}, 267203 (2005).
\bibitem{Lukic} J. Lukic, E. Marinari, O. C. Martin, and S. Sabatini, J. Stat. Mech.: Theory Exp. L10001 (2006).
\bibitem{Ledoussal} P. Le Doussal, Phys. Rev. Lett. {\bf 96}, 235702 (2006).
\bibitem{Rizzo} T. Rizzo and H. Yoshino, Phys. Rev. B {\bf 73}, 064416 (2006).
\bibitem{Katzgraber} H. G. Katzgraber and F. Krzakala, Phys. Rev. Lett. {\bf 98}, 017201 (2007).
\bibitem{Yoshino} H. Yoshino and T. Rizzo, Phys. Rev. B {\bf 77}, 104429 (2008).
\bibitem{Pixley} J. H. Pixley and A. P. Young, Phys Rev B {\bf 78}, 014419 (2008).
\bibitem{Aspelmeier1} T. Aspelmeier, Phys. Rev. Lett. {\bf 100}, 117205 (2008).
\bibitem{Aspelmeier2} T. Aspelmeier, J. Phys. A {\bf 41}, 205005 (2008).
\bibitem{Mora} T. Mora and L. Zdeborova, J. Stat. Phys. {\bf 131}, 1121 (2008).
\bibitem{Aral} N. Aral and A. N. Berker, Phys. Rev. B {\bf 79}, 014434 (2009).
\bibitem{Chen} Q. H. Chen, Phys. Rev. B {\bf 80}, 144420 (2009).
\bibitem{Jorg} T. J\"{o}rg and F. Krzakala, J. Stat. Mech.: Theory Exp. L01001 (2012).
\bibitem{Lima} W. de Lima, G. Camelo-Neto, and S. Coutinho, Phys. Lett. A  {\bf 377}, 2851 (2013).
\bibitem{Katzgraber2} W. Wang, J. Machta, and H. G. Katzgraber, Phys. Rev. B {\bf 92}, 094410 (2015).
\bibitem{MMayor} V. Martin-Mayor and I. Hen, Scientific Repts. {\bf 5}, 15324 (2015).
\bibitem{ZZhu}Z. Zhu, A. J. Ochoa, S. Schnabel, F. Hamze, and H. G. Katzgraber, Phys. Rev. A {\bf 93}, 012317 (2016).
\bibitem{Katzgraber3}W. Wang, J. Machta, and H. G. Katzgraber, Phys. Rev. B {\bf 93}, 224414 (2016).
\bibitem{MMayor2} J. Marshall, V. Martin-Mayor, and I. Hen, Phys. Rev. A {\bf 94}, 012320 (2016).
\bibitem{Fernandez} L. A. Fernandez, E. Marinari, V. Martin-Mayor, G. Parisi, and D. Yllanes, J. Stat. Mech.: Theory Exp., 123301 (2016).


\bibitem{Collet} P. Collet and J.-P. Eckmann, \textit{Iterated Maps on the Interval as
Dynamical Systems} (Birkh\"{a}user, Boston, 1980).
\bibitem{Hilborn} R. C. Hilborn, \textit{Chaos and Nonlinear Dynamics}, 2nd ed. (Oxford
University Press, New York, 2003).

\end{references}
\end{document}